\documentclass[fleqn,usenatbib]{mnras}

\usepackage{newtxtext,newtxmath}

\usepackage[T1]{fontenc}

\DeclareRobustCommand{\VAN}[3]{#2}
\let\VANthebibliography\thebibliography
\def\thebibliography{\DeclareRobustCommand{\VAN}[3]{##3}\VANthebibliography}

\usepackage{graphicx}	
\usepackage{amsmath}	
\usepackage{multirow}   

\newcommand{\msun}[0]{\mathrm{M}_\odot}

\title[Evolution of LYRA dwarfs]{Co-evolution of baryons and dark matter halos of LYRA dwarf galaxies}

\author[J. Sureda et al.]{Joaquin Sureda$^{1,2}$\thanks{E-mail: joaquin.m.sureda@durham.ac.uk},
Shaun T. Brown$^{1,3}$,
Azadeh Fattahi$^{1,3}$, 
Thales Gutcke$^{4}$,
Sownak Bose$^{1}$,
\newauthor
Jessica E. Doppel$^{1,5}$,
Rüdiger Pakmor$^{6}$
\\
$^{1}$Institute for Computational Cosmology, Department of Physics, Durham University, South Road, Durham DH1 3LE, UK\\
$^{2}$European Southern Observatory, Karl-Schwarzschild-Straße 2, 85748 Garching bei München, Germany\\
$^{3}$The Oskar Klein Centre, Department of Physics, Stockholm University, Albanova University Center, 106 91 Stockholm, Sweden\\
$^{4}$Institute for Astronomy, University of Hawaii, 2680 Woodlawn Drive, Honolulu, HI 96822, USA\\
$^{5}$Centre for Extragalactic Astronomy, Department of Physics, Durham University, Durham DH1 3LE, UK\\
$^{6}$Max-Planck-Institut für Astrophysik, Karl-Schwarzschild-Str. 1, D-85748, Garching, Germany
}

\date{Accepted XXX. Received YYY; in original form ZZZ}

\pubyear{2015}

\begin{document}
\label{firstpage}
\pagerange{\pageref{firstpage}--\pageref{lastpage}}
\maketitle

\begin{abstract}
We use the extremely high-resolution ($m_{\rm baryon}=4\rm{M}_\odot$) LYRA cosmological galaxy formation simulations of six dwarf galaxies with $M_{\rm 200c}\sim10^9\rm{M}_\odot$ at $z=0$ to investigate their stellar assembly histories. Based on the age of stars in these galaxies at $z=0$, $40-100\%$ of their stellar mass was formed by the time of reionization, when star formation (SF) abruptly shuts down. Depending on their halo mass evolution, some of these dwarfs reignite SF post-reionization (rejuvenators), while others remain quenched for the rest of cosmic time (reionization relics). However, the stellar mass of the relics can still grow by more than $50\%$ through mergers post-reionisation. We find clear correlations between metallicity distributions of the galaxies and the fraction of stars formed post-reionization ($f_{\rm post-reio}^\star$) such that relics have lower median $\rm [Fe/H]$ with a more prominent low metallicity tail. Moreover, the shape of the galaxies at $z=0$ correlates with their $f_{\rm post-reio}^\star$, with rejuvenators showing more spherical stellar distribution than relics. This difference arises only post-reionization when rejuvenators become rounder with more SF activity. Similarly, the shape of dark matter (DM) halos in the inner regions display more spherical distributions in rejuvenators than in relics. The shape evolution shows that DM haloes in all galaxy formation simulations become rounder in comparison to their collision-less, DM-only counterparts. However, DM haloes of rejuvenators evolve more significantly. We do not find any correlation between SF activity and formation of shallow DM density cores in these galaxies. 
These predictions can be tested using upcoming observational data. In particular, our results indicate that the scatter in the mass-metallicity relation in the low mass regime is correlated with SF histories and the shape of galaxies.

\end{abstract}




\section{Introduction}

Dwarf galaxies are amongst the smallest and most dark matter dominated known structures in the Universe. Such systems have proven powerful laboratories to constrain the nature of dark matter (DM). Their shallow gravitational potentials, and small host DM haloes, make them particularly sensitive to internal feedback processes such as stellar winds, supernovae and AGN feedback \citep{Larson:1974, Saito:1979, Dekel:1986, Stinson:2009, Governato:2010, Sawala:2010, Hopkins:2014, Ceverino:2014, Agertz:2015}.

Observed dwarfs galaxies show a wide diversity in their stellar and gaseous properties. Many bright dwarfs ($M_{\star} = 10^8$--$10^9\rm{M_{\odot}}$) exhibit central thick stellar discs embedded in a diffuse spherical distribution of stars \citep[e.g.][]{Kado-Fong:2020}, while smaller `classical' dwarfs have no clear flattened disc component and instead are elliptical over all radii \citep[e.g.][]{Carlsten:2021}. Dwarf galaxies similarly exhibit a wide range of star formation histories with many bright dwarfs being star forming today \citep[e.g.][]{Geha_12}, even having strong observed gaseous outflows \citep[][]{McQuinn_19, Hamel-Bravo_24,Reichardt-Chu_25}, while fainter dwarfs and satellites of Milky Way mass systems are predominantly quiescent \citep[e.g.][]{Geha_17, Mao_21} and composed of extremely old stellar populations \citep[e.g.][]{Weisz_14}. It is expected for many of these fainter galaxies to have been permanently quenched during the epoch of reionization by the external ionising background \citep{Efstathiou:1992, Dijkstra:2004, Hoeft:2006, Trujillo-Gomez:2015, Benitez:2020}.

To model the dwarf galaxy population many works utilise cosmological hydrodynamic simulations, that self consistently model the joint effects of gravity, hydrodynamics and key astrophysical processes such as star formation and feedback from the very early Universe through to the present day. Large volume simulations offer robust statistics and probe the full range of possible cosmological environments but are only able to resolve dwarf galaxies with modest resolution \citep[e.g.][]{EAGLE,TNG,Colibre}, while dedicated zoom-in simulations of individual dwarfs are able to use significantly higher resolution, with recent generations now directly resolving individual supernovae and massive stars \citep[e.g.][]{Wheeler_19, Agertz:2020, Gutcke:2022-LYRAII, Andersson_25}. The modelling of dwarf galaxies within a cosmological context is an active and ongoing effort within the field, and has proven to be a particular challenge with little convergence in the predicted properties from different simulation efforts \citep[see for example][]{Sales:2022}. However, there are a number of key results that naturally emerge and are consistent with observations. Simulations often exhibit a transition from disc-like, rotationally-supported systems at larger stellar masses to spheroidal, velocity dispersion-supported systems at lower masses \citep[e.g.][]{Pillepich:2019}. Simulated dwarf galaxies display a large variety of star formation histories (SFHs) with larger systems that are star forming today, and smaller galaxies that are quiescent and ceased star formation early in the Universe.

Theoretical studies have shown that cosmic reionization plays a key role in regulating star formation in the early Universe, with many models predicting dwarf galaxies to be directly quenched by the introduction of this ionising background. Moreover, observations of dwarf galaxies in the Local Group show diverse star formation histories with a mass dependence, and there are tentative evidence of the quenching by reionization for the lowest mass satellites of Milky Way \citep{Weisz:2019, Savino:2023, Savino:2025}. The effect of the ionising background on a galaxy's ability to remain star forming can be phrased as a redshift dependent halo mass threshold \citep[e.g.][]{Benitez:2020}, where the mass threshold increases near instantaneously at reionization. In this way the SF histories of dwarf galaxies are expected to be intimately tied to the growth of their DM halo, leading many models to exhibit strong correlations between galaxies' present day stellar component and their DM halo accretion history \citep[e.g.][]{Fitts:2017-FIRE, Matthee_17, Rey:2019}. Many simulations also naturally produce the observed transition in the observed stellar morphologies with many bright dwarfs forming thick discs, while classical and ultra faints exhibiting spheroidal stellar distributions, becoming more spherical towards lower masses \citep[e.g.][]{Pillepich:2019, Liao:2019, Jiang:2019}.


While the host DM halo and its growth plays a crucial role in a galaxy's evolution, there are important back reactions from the galaxy and its evolution on the host DM halo. One of the most striking ways in which this can manifest is the formation of DM cores. In the absence of any baryonic processes DM haloes are well approximated by an NFW density profile \cite{Navarro:1996}, with an inner `cusp' ($\rho_\mathrm{DM} \propto r^{-1}$). When baryonic processes are included, multiple simulations exhibit haloes with cored central density profiles \citep[e.g.][]{Di-Cintio_14,Tollet_16,Maccio_17}. Through feedback, gas can be driven from the centre of galaxies and in the processes vary the gravitational potential over short time scales, efficiently transferring energy to the DM halo and leading to core formation \citep{Read:2005,  Pontzen:2012}. While it is well established that feedback \textit{can} form cores, many cosmological galaxy formation simulations do not form any DM cores at any time, in some cases even resulting in a contraction of the halo \citep[e.g][]{Bose:2019, Cautun_20}. Indeed, first analysis of a LYRA dwarf did not show the presence of a core. Studies suggest that efficient baryon-driven core formation requires two key criteria: (i) the gas should be able to dominate the central potential prior to it being driven as an outflow, (ii) that there are periodic and ongoing feedback events that continually transfer energy to the DM halo and prevent a cusp reforming \citep[e.g.][]{Pontzen:2012, Benitez:2019}. The discrepancies between current simulations suggest that core formation, or the lack of, is subject to subtle choices in how these galaxy formation models, and their feedback prescriptions, are implemented, with no current consensus. This is particularly relevant when comparing to observations of dwarf galaxies. While some observations of nearby dwarf galaxies \citep{deBlok:2008, Oh:2011} point to shallow dark matter density profiles, other works discuss the challenges to successfully infer the dark matter content in the inner region of galaxies, even finding contradictory results to previous studies when some corrections are taken into account \citep{Strigari:2017, Genina:2018, Harvey:2018}. Therefore, although the cored density profiles can be explained, the focus is now on understanding the large diversity of rotation curves in this mass regime \citep{Oman:2015, Oman:2019, Santos-Santos:2020, Sales:2022}.

In addition to baryonic processes affecting the radial distribution of DM, they also have a significant effect on the angular distribution, and shape, of DM haloes. Many studies have measured how halo shape depends on the mass, redshift and assumed cosmology in the absence of baryons, concluding that larger haloes are the most ellipsoidal with smaller haloes becoming closer to spherical \citep[e.g.][]{Allgood:2006, Tissera:2010, Vera-Ciro:2011, Prada:2019}. The inclusion of baryons generally leads to individual DM haloes becoming significantly more spherical, even when the baryons are distributed axisymmetrically \citep[e.g.][]{Chua_19,Cataldi:2021}. Recently, the EDGE collaboration \citep{Agertz:2020} carried out a detailed study of the halo shapes in the dwarf galaxy regime \citep{Orkney:2023}. They find that this transition towards rounder halos when including baryons does not always hold true for dwarf galaxies, and depends strongly on the gas content of their dwarfs.

In this work we aim to further understand the complex relation between the host DM halo and its galaxy within the dwarf galaxy regime. To do this we use the LYRA suite of extremely high-resolution hydrodynamical cosmological zoom-in simulations of dwarf galaxies \citep{Gutcke:2021-LYRAI,Gutcke:2022-LYRAII,Gutcke:2022-LYRAIII}. Their unique resolution, $4 \rm{M_{\odot}}$ per baryonic resolution element, allows us to follow the evolution of DM, gas and stars in exquisite detail from the very early universe at the birth of these galaxies to the present day. The paper is organised as follows. In Section \ref{sec:methods} we briefly introduce the main aspects of the LYRA model, the sample of galaxies, and describe the methods used in this work. Our main results are presented in Section \ref{sec:results}, focusing on characterizing the galaxies based on their SFHs and then investigating differences among the galaxies themselves. Finally, the main conclusions of this work are highlighted in Section \ref{sec:conclusions}.

\section{Methods}\label{sec:methods}

\subsection{LYRA model}\label{sec: lyra model}

In this study, we analyse a sample of six dwarf galaxies, which were simulated using the zoom-in technique \citep{Jenkins:2013}  in a $\Lambda$CDM universe, utilising the cosmological hydrodynamical code \textsc{arepo} \citep{Springel:2001, Pakmor:2016, Weinberg:2020} and the LYRA galaxy formation model \cite{Gutcke:2021-LYRAI, Gutcke:2022-LYRAII}. The sample includes 5 galaxies presented first in \cite{Gutcke:2022-LYRAIII}, labelled from A to E. In addition, a new halo is included in the analysis, labelled as halo F and briefly presented in \citet{Gutcke:2024}. 

With a baryonic resolution of $4~\msun$, the LYRA model includes a resolved multiphase interstellar medium (ISM) which cools down to $10$ K. The formation and subsequent evolution of individual stars and completely resolved supernovae blastwaves. In addition, Pop III star metal enrichment is included as part of a subgrid model to enhance the metallicities at high redshift \citep{Gutcke:2022-LYRAII}. The non-equilibrium cooling rates for neutral hydrogen and helium are computed on the fly, while metal lines and low temperature ($\mathrm{H}_2$) cooling rates are taken from \cite{Ploeckinger:2020}, which assumes equilibrium ionisation rates.  
The ultra-violet background (UVB) is implemented as a homogenous and isotropic heating term following \citet{Faucher:2020}, including the self-shielding of gas as presented in \citet{Rahmati:2013}. With this prescription, hydrogen reionization occurs at $z=7.8$. Finally, concerning stellar formation modelling, LYRA follows a Schmidt relation \citep{Schmidt:1959} for the star formation. Specifically, the gas cells with $T<100$K can form stars, with efficiency increasing as a function of the gas density, starting at $n_\mathrm{H}>10^3 \mathrm{cm}^{-3}$ and reaching a $100\%$ efficiency at $n_\mathrm{H}>10^4 \mathrm{cm}^{-3}$. The stars are sampled from a Kroupa IMF \citep{Kroupa:2001}, where stars above the minimum mass $M_{*,\min}=4\,\msun$ are resolved and sampled individually, and below that this threshold the star particles represent the integrated population.

The zoom-in initial conditions are drawn from the 100 Mpc DMO box of the EAGLE project \citep{Schaye:2015}, with cosmological parameters consistent with Planck 2013 results \citep{Planck:2014}, i.e. a flat universe with $h=0.6777$, $\Omega_{\rm m}=0.307$, $\Omega_{\rm bar}=0.048$, $\Omega_{\rm \Lambda}=0.693$, $\sigma_8=0.8288$. The DM mass resolution within the high-resolution region is $\sim80 \msun$. In addition to the \texttt{Subfind} \citep{Springel:2001} halo catalogue, for each of the runs, we construct merger trees using the \texttt{D-halos} tree code \citep{Jiang:2014} to be able to follow more accurately the growth histories of the simulated halos. This allows for the proper identification of significant mergers that might have an impact on the overall properties of these galaxies.

\begin{figure}
    \centering
    \includegraphics[width=\columnwidth]{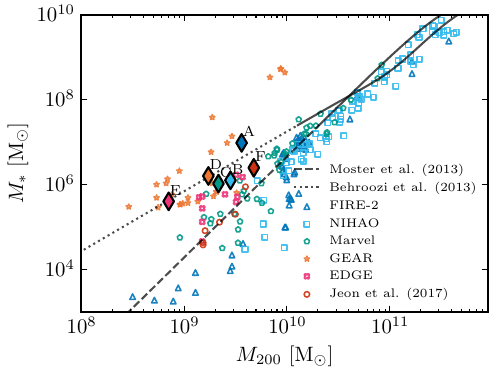}
    \caption{Galaxy stellar mass vs dark matter halo mass relation. The diamonds show the six LYRA galaxies in this relation. Other symbols show isolated galaxies from different simulations (data compiled in \citet{Sales:2022}). In addition, the solid lines illustrate abundance matching results with the dashed and dotted lines corresponding to extrapolations on the low mass regime.}
    \label{fig:Stellar-to-haloMass}
\end{figure}

Fig. \ref{fig:Stellar-to-haloMass} presents the stellar mass-halo mass\footnote{Halo properties are measured within $R_{200\rm c}$, i.e. the radius where the mean enclosed density is 200 times the critical density of the Universe.} (SMHM) relation for the six galaxies in the sample, indicated by the diamonds. The LYRA halos, which span a halo mass range of $7\times 10^8 - 5\times 10^9 \,\msun$, host galaxies with stellar masses in the range $4\times 10^5 - 1\times 10^7 \,\msun$ at $z=0$. In the same figure, there are samples of other simulated dwarf galaxies from other projects, namely, FIRE-2 \citep{Hopkins:2018}, NIHAO \citep{Wang:2015}, MARVEL-ous, \citep{Munshi:2021}, GEAR \citep{Revaz:2018}, EDGE \citep{Agertz:2020, Rey:2022} and \citet{Jeon:2017}. At halo masses probed by our LYRA galaxies, there is a noticeable difference in predictions from various models. LYRA galaxies lie towards the upper end of this scatter. They are comparable with GEAR dwarfs and upper end of the EDGE sample, as well as extrapolation of the abundance matching relation from \citet{Behroozi:2013}. We note that the scatter in stellar mass of EDGE dwarfs is due to the assembly history of halos by design (using their genetically modified initial conditions), and higher stellar masses correspond to halos which assembled earlier \citep{Rey:2019}.

In terms of mass resolution, EDGE is the closest one to LYRA, with a baryonic resolution of $~20 \msun$, followed by FIRE-2 with a baryonic resolution of $30 \msun$, which are the lower stellar mass points in Fig. \ref{fig:Stellar-to-haloMass}. Nevertheless, these do not agree with each other, emphasising the need for better understanding of galaxy formation in this mass regime combined with observational constraints.

\subsection{Merger identification}\label{sec: merger id}

To assess the impact of \textit{significant} mergers on the evolution of LYRA galaxies we follow the procedure below to compute the merger ratios and merger times. We first identify the main progenitor of our six galaxies throughout time, at least back to $z=10$. The reason for this redshift cut off is that, both in practice and in principle, identifying the main progenitor at very high redshift is not straightforward, due to the large rates of mergers, many of which are approximately equal in mass.
We mainly follow the main branch of the merger tree to identify the main progenitors. Occasionally, in particular during mergers, the main progenitor is not identified properly by the merger tree, which becomes evident by a big jump in mass at the time. Therefore, we manually fix these instances. Once the main progenitor is known at each snapshot, we identify all the possible progenitors at the previous snapshot and track them back in time. We define the infall time $t_\mathrm{infall}$ when the progenitor crosses the main halo $R_{200}$ for the first time. We then measure the mass ratio of each progenitor one crossing time $t_\mathrm{cross}$ before the infall time, where the crossing time is defined by

\begin{equation}
    t_\mathrm{cross} = \frac{2R_{200}}{\sigma_v},
\end{equation}
with $R_{200}$ and $\sigma_v$ correspond to the virial radius and the velocity dispersion of the main halo, respectively. This allows us to measure the mass of the infalling halo before it starts losing mass due to the interaction with the main halo. 

Finally, we define significant mergers as those with mass ratios above $1\colon10$ and only show these in the following analysis. For these mergers, we inspect their evolution in more detail to ensure we measure the mass ratio at a reasonable time. The merger mass ratio is defined as the total (dark plus baryonic matter) mass ratio.

\subsection{Shape determination}\label{sec: shape-algorithm}

We use the method described in \citet{Allgood:2006} to quantify the ellipsoidal shape of the galaxies and their halos. This can also be viewed as a proxy for the morphology of the baryonic components. This method assumes the particle distribution to be approximated to an ellipsoid and uses the reduced inertia tensor,
\begin{equation}
    \Bar{I}_{ij} = \sum_{n} \frac{m_n x_{i,n} x_{j,n}}{r_n^2},
\end{equation}
where the summation is over the number of particles, n,  and $r_n = \sqrt{x_0^2 + x_1^2/q^2 + x_2^2/s^2}$ is the elliptical radius of each particle where the coordinate system is aligned with the principal axes of the ellipsoid with $q$ and $s$ representing the intermediate-to-major and minor-to-major axis ratios of the ellipsoid, respectively. 
The eigenvectors of the inertia tensor define the orientation of the ellipsoid, while the length of the major axes of the ellipsoid, $a\geq b\geq c$, are given by the square root of the eigenvalues, $\lambda_0 \geq \lambda_1 \geq \lambda_2$ with axis ratios defined as $q=b/a=\sqrt{\lambda_1/\lambda_0}$ and $s=c/a=\sqrt{\lambda_2/\lambda_0}$. 
In practice, the shape of the ellipsoid is found iteratively and in each iteration the length of the major axis of the ellipsoid, a, is kept fixed to the initial value for the iteration.
We start by selecting particles enclosed in a sphere of fixed radius $r_\mathrm{init}$ (i.e. an ellipsoid with $a=r_\mathrm{init}$ and $q=s=1$ ) from which the inertia tensor is computed. From the corresponding eigenvalues and eigenvectors, a new ellipsoid is drawn from which a new set of particles is selected. This process is repeated until a convergence criterion is reached. In this work, we say the process has converged if the relative difference of the axis ratios ($q$ and $s$) between the iterations is within $0.1\%$, with the condition of having a minimum of 1000 particles to compute the inertia tensor. Finally, we use the resulting axis ratios to compute the triaxiality parameter 

\begin{equation}
    T = \frac{1-q^2}{1-s^2},
    \label{eq: Triaxiality}
\end{equation}
which serves as a summary parameter that captures the overall particle distribution in a single value. Using this parameter, the shape of a component is usually defined as \textit{Prolate} for $0.7 < T \leq 1$, \textit{Triaxial} for $0.3 \leq T < 0.7$ and \textit{Oblate} for $0 \leq T < 0.3$. For instance, a disc-like morphology appear as highly oblate with very small triaxiality parameter. See Table \ref{tab:morphology-summary} for a summary of these definitions. While it is an useful parameter to describe the particle distribution, it does not retain the full information of the ellipsoid. Therefore, we will usually show both, the axes ratios and the triaxiality parameter when discussing the shapes of the galaxies and halos.

\begin{table}
    \centering
    \begin{tabular}{cp{0.2\linewidth}p{0.5\linewidth}}
        \hline
        Label & $q,s$ $;T$ & Description\\ \hline
        Spherical &  $1 \gtrsim q \gtrsim s$ \newline $0 < T \leq 1$ & Ellipsoid with nearly equal axes, not adequately described by triaxiality due to its full value range.\\
        Prolate &  $1 \gg q \geq s$ \newline $0.7 < T \leq 1$ & Elongated ellipsoid stretched along its major axis, resembling the shape of a rugby ball.\\
        Triaxial & $1 > q > s$ \newline $0.3 \leq T < 0.7$ &  Ellipsoid with three distinct principal axes of different lengths, resulting in an asymmetric, stretched shape.\\
        Oblate & $1 \geq q \gg s$ \newline $0 \leq T < 0.3$ & Flattened ellipsoid compressed along its minor axis, resembling the shape of a squished sphere or a disc.\\
        \hline
    \end{tabular}
    \caption{Summary of the morphological definitions based on the inertia tensor method.}
    \label{tab:morphology-summary}
\end{table}

\section{Results}\label{sec:results}

The original sample of five LYRA galaxies was introduced in \citet{Gutcke:2022-LYRAIII}, with an extensive study on the SFHs of the dwarfs. They focused on studying the reasons behind the diverse SFH of the sample, in particular investigating why some of the galaxies have extended SFHs after reionization. Overall, the fate of dwarf galaxy in the mass regime relevant to this study, is determined by the host halo assembly history, and in particular halo mass at reionization, as suggested by \citet[][BL20 hereafter]{Benitez:2020}; i.e. reionization suppressed SFHs in halos with masses below $2\times10^8 \msun$ (at $z=7.7$). Nevertheless, close to this empirical threshold of star formation, the detailed properties the of the galaxies and their host halo can significantly influence their evolution, as shown by \citet{Gutcke:2022-LYRAIII}. 
For instance, the star formation activity is directly related to the capacity of the halo to retain cold gas shortly after reionization. Therefore, although this is strongly dependent on the halo mass at reionization, it will also be affected by the details of each individual halo at this time.

Aside from the direct influence of diverse SFHs, one might expect secondary observable differences among the galaxies that continue forming stars, and the ones that remain quenched after reionization. In this work, we focus on analysing the consequences of these different evolutionary paths. Additionally, we introduce a new halo in the analysis, and we use the recently run merger trees to ensure the correct tracking of the main halo across the simulations. Then, we start by highlighting the different growth histories of the simulated galaxies.

\begin{table*}
    \centering
    \begin{tabular}{cccccccccccc}
    \hline
    Halo & $M_\star$ & $M_\mathrm{200c}$ &  $r^{\star}_{1/2}$ & $R_\mathrm{200c}$ & $z_{90}$ & $\tau_{90}$ & $z_\mathrm{rej}$ & $t_\mathrm{rej}$ & $M_{\mathrm{200c}}(z_\mathrm{reio})$ & $f^{\star}_\mathrm{post-reio}$ & $f_\mathrm{acc}$  \\
     &  [$10^6 \, \mathrm{M}_{\sun}$] & [$10^9 \, \mathrm{M}_{\sun}$] &  [pc] & [kpc] & & [Gyr] &  & [Gyr] & [$10^8 \, \mathrm{M}_{\sun}$] & & \\
    \hline
     &\multicolumn{10}{c}{\textbf{Rejuvenators}}\\
    A & $9.58$ & $3.62$ & $330.7$ & $32.38$ &$2.20$ & $3.00$ & $5.15$ & $1.14$ & $1.04$ & $0.59$& $0.20$  \\
    F & $2.48$ & $4.76$ & $890.8$ & $35.45$ &$0.87$ & $6.44$ & $3.44$ & $1.85$ & $0.33$ & $0.11$& $0.22$  \\
    D & $1.59$ & $1.72$ & $119.6$ & $25.27$ &$4.95$ & $1.19$ & $5.01$ & $1.18$ & $1.95$ & $0.23$& $0.39$  \\
     &\multicolumn{10}{c}{\textbf{Relics}}\\
    B & $1.26$ & $2.82$ & $483.2$ & $29.79$ &$8.76$ & $0.57$ & $4.58$ & $1.32$ & $0.62$ & $0.02$& $0.39$  \\
    C & $1.05$ & $2.16$ & $878.4$ & $27.24$ &$9.25$ & $0.53$ & -      & -      & $0.25$ & $0$   & $0.33$  \\
    E & $0.40$ & $0.71$ & $641.9$ & $18.45$ &$8.92$ & $0.56$ & -      & -      & $0.19$ & $0$   & $0.15$  \\
    \hline
    \end{tabular}
    \caption{Summary of relevant parameters: Stellar mass, $M_{200\mathrm{c}}$, stellar $r_{1/2}$ and $R_{200\mathrm{c}}$ at $z=0$; redshift and time for the assembly of $90\%$ of the stellar mass, redshift and time for the first SF episode after reionization (rejuvenation time), $M_{200\mathrm{c}}$ at reionization, fraction of stellar mass formed after reionization, and the fraction of stellar mass accreted post-reionization from the most relevant mergers in the main branch.}
    \label{tab: dwarf info}
\end{table*}

\subsection{Assembly histories of the dwarfs}
\label{sec:assemblyhistory}

\begin{figure*}
    \centering
    \includegraphics[width=2\columnwidth]{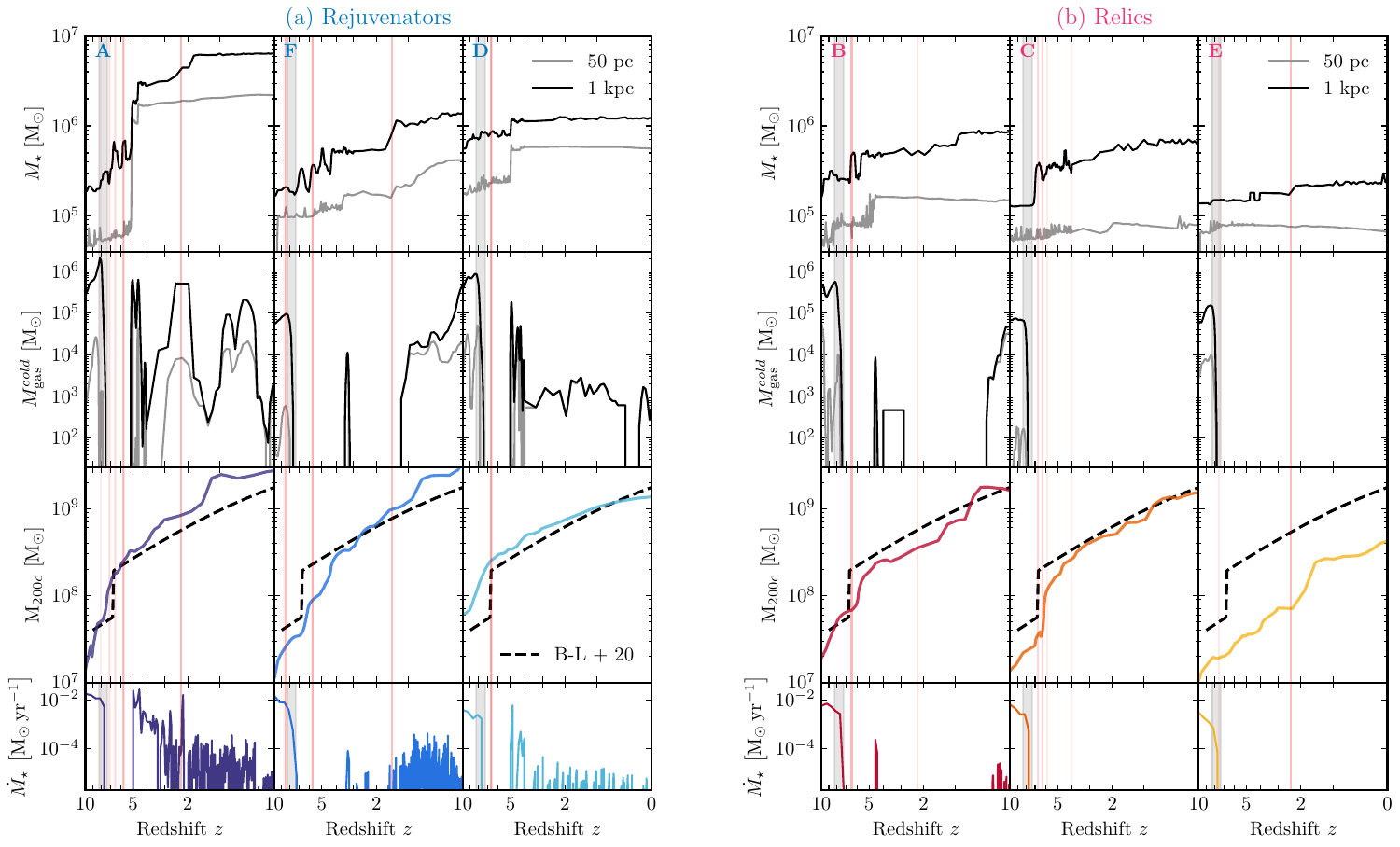}
    \caption{Stellar (top) and cold gas (mid) mass growth vs z at different radius, where cold gas refers to gas cells with temperature $T_\mathrm{gas}<10^3 \,\mathrm{K}$. The bottom panel shows the SFH for each halo in 50 Myr bins. The left panels correspond to rejuvenated galaxies whereas the right ones are reionization relics. The grey vertical band illustrates reionization while the red translucent vertical lines indicate the corresponding significant mergers for each halo. Fainter merger lines correspond to smaller merger mass ratios. We see overall significant differences between the relics and rejuvenated dwarf galaxies, in terms of their SFHs and their cold gas content, illustrating our categorization of each galaxy in the sample. }
    \label{fig:Mass growth and SFH}
\end{figure*}

In this section we examine further the stellar assembly history of the dwarf galaxies in the LYRA sample, alongside the evolution of their gaseous component, and the role of mergers. In Fig. \ref{fig:Mass growth and SFH}, we show the evolution of baryons in our sample (following the main progenitor in time), where each column corresponds to an individual galaxy. The stellar and cold gas mass growth, measured inside two different radii (see the legend) are shown in the first two rows, respectively. The cold gas here refers to gas cells with temperature $T_\mathrm{gas}<10^3 \,\mathrm{K}$. For completeness, we show the halo assembly history in the third row, along with the SF threshold derived in \citet{Benitez:2020}. Finally, the bottom row of this figure highlights the star formation rate as a function of redshift (previously shown in \citet{Gutcke:2022-LYRAIII} for 5 of these galaxies, except Halo F). The SFRs are computed by binning the ages of all the stars associated with the central galaxy (defined by \texttt{Subfind}) at $z=0$, thus, it does not distinguish between accreted vs in-situ star formation. 

All the galaxies are significantly impacted by cosmic reionization, indicated by the grey vertical band at $z\sim8$, as shown by the sharp drop in SFR and their cold gas content at this time. Nevertheless, some galaxies are able to reignite their star formation sometime after reionization. We group galaxies that reignite after reionization in the left side of Fig. \ref{fig:Mass growth and SFH} and refer to them as \textit{rejuvenators}. The presence of cold gas and non-zero SFRs for an extended period after reionization characterises this group. The rest of the sample that remain quenched (arranged in the right side) are referred to as \textit{reionization relics}. We include a summary of the relevant parameters of all relic and rejuvenated halos in Table \ref{tab: dwarf info}, sorted by decreasing stellar mass.
We only classify dwarfs as rejuvenators if they present a sustained star formation activity after reionization. Hence, we are not considering galaxies with only sporadic star formation episodes as rejuvenated (e.g. Halo B). 
In order to give a more quantitative criteria for separating the two subgroups, we also look at the redshift (or time) when 90\% of the stellar mass is assembled, $z_{90}$ (or $t_{90}$). Equivalently, we compute the stellar mass fraction formed post-reionization, $f^{\star}_\mathrm{post-reio}$. These values are quoted in Table \ref{tab: dwarf info} and give a clear picture of the dichotomy in our sample. For example, we can see a clear gap in $z_{90}$ of the two groups with relics having $z_{90} > z_\mathrm{reio}$. \footnote{We note that halo F is an marginal case with $f_\mathrm{post-reio}^{\star}=0.11$. Nevertheless, due to its sustained star formation since $z=2$, we classify it as a rejuvenator.}  It is worth emphasizing that the halo mass at reionization alone (also given in Table \ref{tab: dwarf info}) is not a reliable predictor for whether a galaxy will be a rejuvenator or not. Hence, the total assembly history of the halo should be considered.

We see in Fig. \ref{fig:Mass growth and SFH} that the SF activity of galaxies closely trace the presence of cold gas content in these dwarf galaxies.    
However, higher star formation rates result in the drop in cold gas mass as can be seen clearly in Halo A and D after $z\sim 5$. The semi-empirical model of \citet{Benitez:2020} predicts that halo mass at a given time (halo assembly history) determines the ability of the gas in the halos to cool down and form stars, and \cite{Gutcke:2022-LYRAIII} shows that LYRA galaxies roughly follow that prediction (see also third row panels of Fig. \ref{fig:Mass growth and SFH}). 
However, as mentioned before, close to the BL20 star formation threshold, the particular evolution of each halo can affect this prediction. For example, based on the prediction of BL20, Halo D should have survived reionization, but it loses all its cold gas and stops forming stars at reionization until $z\sim5$. On the other hand, Halo B is able to cool down its gas and have a short episode of star formation around $z\sim 5$ despite being below the star formation threshold of BL20. Halo F is a curious rejuvenator with a long period of having almost no cold gas, after reionization until $z\sim 1$. This is linked to its delayed halo mass growth. This halo is amongst the lowest mass halos (relics) prior to $z\sim 6$, but it grows more rapidly at more recent times due to a merger (more details below), and it will eventually become the most massive halo of our sample by $z=0$. The gas can only cool down in this halo at lower redshifts, $z<1$. This rather unusual assembly history leaves some imprint in the observable properties of this halo as we see in the next section.

Next, we turn to the stellar mass growth of the dwarfs which is directly linked to the SFHs.  We measure the enclosed stellar mass at two radii, an inner and an outer radius (50 pc and 1 kpc respectively), shown in each top sub panel in Fig. \ref{fig:Mass growth and SFH}. These radii were chosen to trace the innermost region of the galaxy, as well as most of the galaxy content, respectively. Note that the half stellar mass radii of the galaxies are between $100-900$ pc (see Table \ref{tab: dwarf info}). 
There are a number of features worth noting in the stellar mass growth. Firstly, we note that the first peak of star formation after reionization, which typically happens around $z\sim 5$ for these galaxies, causes a sharp increase in the stellar mass in the very central region, while the fractional increase is less prominent when considering the whole galaxy. This points to a centrally concentrated star formation. For example, the stellar mass in the central 50 pc region of halo A increases by more than a factor 10 (mass increase of $1.04\times10^6 \msun$) at $z=5$, and the overall stellar mass increase inside 1 kpc is $1.36\times10^6 \msun$, indicating that all of the newly formed stars are concentrated in the centre. Similar behaviour is seen halos D and F, and even B, but to a lesser degree.   
This concentration of stellar mass would explain the central stellar overdensity reported by \citet{Gutcke:2022-LYRAIII}, suggesting that this starburst marks the moment where this overdensity arises. 

\subsubsection{The role of mergers}

Another feature apparent in the stellar mass growth curves in Fig. \ref{fig:Mass growth and SFH} is the sudden increase in the stellar mass without any star formation activity (or very low SFRs). For example, Halo F has a significant increase in the overall stellar mass at $z\sim1$ before having any cold gas for star formation (the very low SFR can not explain the increase the stellar mass); similarly Halo E has an increase in the overall stellar mass at $z\sim2$ without any star formation activity. Such instances are due to mergers, as shown by the vertical red lines, which mark significant mergers (as defined in Sec. \ref{sec: merger id}). The lines are plotted at $t_\mathrm{infall}$, the moment where the merging dwarfs cross the $R_{200}$ of the main halo for the first time; and the lines' opacities indicate the merger mass ratio, i.e. more massive mergers are drawn as opaque lines, whereas the smaller mergers are drawn fainter. In some instances, there is a short delay between $t_\mathrm{infall}$ and the increase in stellar mass, which is due to the time difference between crossing $R_{200}$ of the main halo and merging with the central galaxy. 

By examining all the galaxies individually, we see the stellar mass growth due to mergers. The last column in Table \ref{tab: dwarf info} gives the fraction of stellar mass at $z=0$ contributed by significant mergers occurring post reionization, $f_\mathrm{acc}$. In all galaxies, the mergers increase the overall stellar mass within $<1$ kpc, but have very little effect in the stellar mass in the central regions, $<50$ pc. This is opposite to in-situ star formation which increases the stellar mass in the central region, as discussed earlier. This difference can be understood by the fact that debris from mergers are deposited on higher energy orbits (``stellar halo-like''), linked to the orbit of the merging objects, whereas the in-situ star formation happens in the dense gas in the centre.  

It is worth highlighting that relic galaxies, which have almost no star formation activity, show a steady growth of the overall stellar mass after reionization which can be explained by mergers and accretion of smaller systems. All relic galaxies grow in stellar mass by a factor of two after reionization \footnote{Note that we only mark mergers with mass ratios larger than $1\colon10$ in this figure.}. These findings are directly linked to the existence of extended stellar components (stellar halos) around these dwarfs. This is an interesting topic in its own right, and it will be the subject of our future works.

We notice that the rejuvenated galaxies tend to have a higher number of significant mergers after reionization, and there is a connection between (some of) the mergers and star formation activity. For example, Halo F displays an increase in the cold gas content shortly after the merger at $z\sim 1.5$, where the galaxy suddenly ignites a sustained period of star formation. We argue that this merger not only brings some gas with it but also helps to cool down some of the available gas currently present in the galaxy that will eventually end up forming stars \citep[see][for discussions on starbursts due to dwarf-dwarf mergers]{Lelli:2014, Lahen:2020, Zhang:2020b, Hislop:2022,Nidhi:2022}. It is important to note that although the halo mass surpassed the threshold required for gas to collapse \citep{Benitez:2020}, it is not until this merger that the halo reignites its sustained star formation. Therefore we argue that mergers can play an important role for galaxies close to this threshold in determining the future star formation history. We stress that this is an argument based on the timing of the merger and the subsequent increase in the cold gas content of the galaxy after that event. We cannot trace back the origin of the gas because LYRA is run without tracer particles for the gas.

Finally, the conditions and events shaping the SFHs of the galaxies are expected to leave imprints on different galaxy properties. These may become apparent when comparing the two categories of galaxies, as the individual conditions (e.g., gas content, DM distribution) differ.

\subsection{Differences between relics and rejuvenated dwarfs}
\label{sec:relicsvsrejuv}

\begin{figure}
    \centering
    \includegraphics[width=\linewidth]{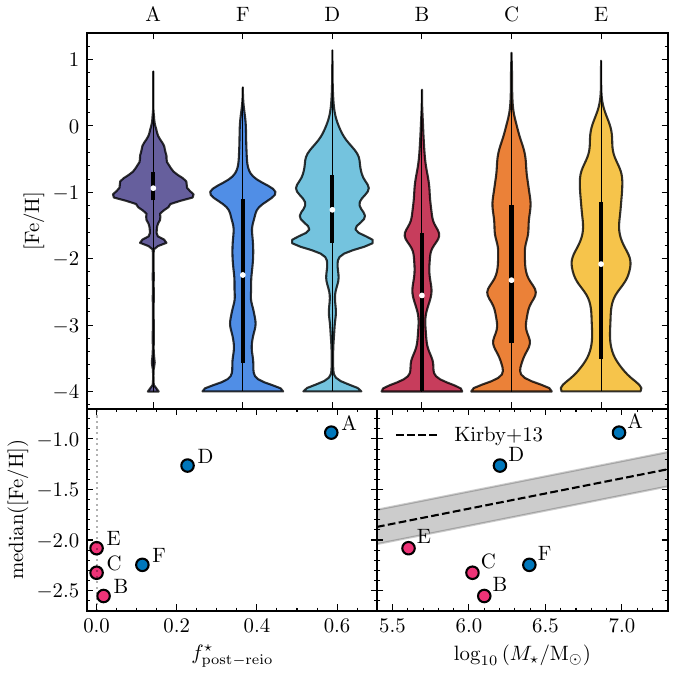}
    \caption{\textit{Top:} Stellar metallicity distribution for each LYRA halo. The dot highlights the median of the distribution with the thicker line showing the interquartile range. \textit{Bottom:} The median stellar metallicity against the mass fraction of stars formed post-reionization, $f^{\star}_\mathrm{post-reio}$ (left), and against the total stellar mass (right). On the right panel we include a mass-metallicity relation for dwarf galaxies from \citet{Kirby:2013}. Relic galaxies show in general more extended metallicity distributions with smaller median values at smaller post-reionization mass fractions.} 
    \label{fig:Metallicity}
\end{figure}

In this section we look into metallicity distributions and the shape of the baryonic components to find a connection with the distinct star formation histories.
One of the first things that comes to mind is the difference in the metallicity distributions at $ z=0$ due to the difference in the SFHs. In the top panel of Fig. \ref{fig:Metallicity}, we show the metallicity distribution of the stars within stellar half-mass radius, $r_{1/2}$, for all galaxies, excluding satellites from the sample. The median of the distribution is highlighted as a white circle and the interquartile range is marked with a thicker line. For each galaxy, the distribution is normalised to its maximum, therefore this figure only illustrates the differences in the shapes of the distributions. We first note that there is a prominent tail of stars with $\left[\mathrm{Fe}/\mathrm{H}\right] = -4$ in all Halos. This is due to the presence of stars at the metallicity floor, as defined by the LYRA model \citep[see][]{Gutcke:2022-LYRAII}. 

The metal-poor tail of the distributions is less prominent in the rejuvenated dwarfs due to their extended star formation that allows for the more metal-rich populations to form and dominate the distribution. In contrast, the relics show a more extended distribution, highlighting the relative prevalence of the metal-poor stellar populations in these galaxies. In particular, Halo C and E, show very similar distributions due to the fact that they only form stars before reionization. Halo B shows a small peak towards slightly higher metallicities, due to the small post-reionization SF episode experienced by this halo. Nevertheless, the median metallicities are consistently smaller for relics than for rejuvenated dwarfs. This is seen more clearly in the bottom left panel of Fig. \ref{fig:Metallicity} showing the median metallicity of the galaxies against the fraction of stellar mass formed after reionization, $f^{\star}_\mathrm{post-reio}$. Here rejuvenated galaxies appear in a blue shade and relics as a magenta colour. 

Aside from the dichotomy in the median metallicity of the rejuvenators and relics, Fig. \ref{fig:Metallicity} highlights a correlation between metallicities and the fraction of stars formed after reionization, $f^{\star}_\mathrm{post-reio}$, which is a proxy for SFH. Relic galaxies with close to zero star formation post-reionization, have  $f^{\star}_\mathrm{post-reio} \simeq 0$, and are grouped in the low metallicity region of the figure. In contrast, rejuvenators have higher fractions and can reach higher median metallicities. The correlation seen in this figure is a result of the metal enrichment history of the galaxies.  For completeness, we show the mass-metallicity relation for the LYRA sample in the bottom left panel (Fig. \ref{fig:Metallicity}), together with the mass-metallicity relation from \citet{Kirby:2013} (black dashed line and shaded region for the scatter) obtained from a sample of Local Group dwarf galaxies. While LYRA galaxies exhibit a slight trend with stellar mass, they show a large scatter compared to the observations. In particular, we observe that relic galaxies always fall under the relation, whereas two of the rejuvenated galaxies are well above the relation. These results suggests that the scatter in the mass-metallicity relation of dwarfs galaxies is likely correlated with their star formation histories.

We notice that the rejuvenator Halo F is amongst the relics in this plane. As discussed in the previous section, some aspects of the evolution of this halo are similar to relics, namely the relatively low mass at high redshift and long period of no star formation activity post-reionization. This galaxy restarts star formation only around $z\sim1$ followed by a significant period of continuous star formation. These properties of Halo F highlight the lack of a sharp definition among these two categories of dwarfs, whereas the parameter $f^{\star}_\mathrm{post-reio}$ can capture the differences better. For easier presentation, we will show our results in terms of relics and rejuvenators but we will include the information of $f^{\star}_\mathrm{post-reio}$ where relevant.

\begin{figure}
    \centering
    \includegraphics[width=\columnwidth]{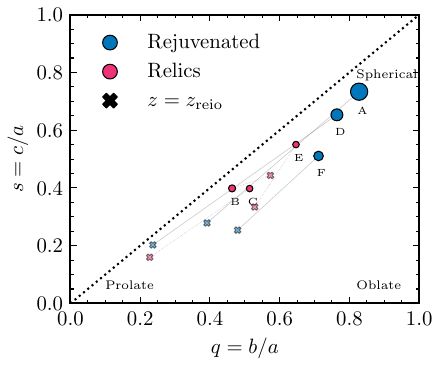}
    \caption{Ellipsoidal axis ratios for the stellar component of the LYRA dwarf galaxies at $z=0$, measured enclosed to the stellar half mass radius, $r_{1/2}$ (see Sec. \ref{sec: shape-algorithm} for the details). The colour of the symbols represents their SFH category, rejuvenator (blue) or relic (red), and the size of each symbol is proportional to the fraction of stellar mass formed after reionization. This figure illustrates that SFH activity of LYRA dwarf galaxies are correlated with their shape.} 
    \label{fig: Stellar Axis ratios}
\end{figure}

Next, we focus on the structural parameters of the galaxies. We investigate this by computing the ellipsoidal shapes using the procedure described in Sec. \ref{sec: shape-algorithm}. In Fig. \ref{fig: Stellar Axis ratios} we present the axis ratios, $q$ and $s$, measured for stars inside the stellar half mass radius, $r_{1/2}$ of LYRA dwarfs. The axis ratios are measured at $z=0$ (circles), and are connected with thin lines to their corresponding shape at the end of reionization (crosses). These are grouped into rejuvenators (blue symbols) and relics (magenta symbols). At $z=0$ (circles) a clear separation between the two groups is observed in this plane. For the rejuvenated galaxies, the axis ratios are larger, with values around $ 0.7\lesssim  q \lesssim 0.9$ and $0.5 \lesssim s \lesssim 0.8$, indicating a more spherical stellar distribution, compared to relics with axis ratios in the range $ 0.45\lesssim  q \lesssim 0.65$ and $0.4 \lesssim s \lesssim 0.6$. Conversely, at the end of reionization (crosses), galaxies appear less spherical and are characterised by a rather prolate morphology, without a clear separation between relics and rejuvenators. More importantly, the difference between the shapes at the end of reionization and at $z=0$ is more pronounced for rejuvenated galaxies, evolving significantly from their shape at reionization, in contrast with relic galaxies.

Moreover, the shape of the galaxies shows a clear trend with the fraction of stars formed post-reionization, $M_\star (> z_{\rm re})/M_\star({z_0})$, indicated by the size of the circles in Fig. \ref{fig: Stellar Axis ratios}. However, we do not see a clear trend with stellar mass of the objects. Note that even if galaxies B, C and D have very similar stellar masses (see Table \ref{tab: dwarf info}), they have clearly differentiated shapes. This illustrates that for LYRA galaxies, the stellar mass alone is not a good predictor for the sphericity of the galaxy and one would need to know the detailed star formation history to better infer the roundness of the stellar component. While these trends are based on a small sample, they nonetheless present a testable prediction connecting SFH and morphology in dwarf galaxies. It is worth noting that recent studies are pushing the boundaries of the inference of the intrinsic morphologies of observed dwarf galaxies \citep{Kado-Fong:2020,Kado-Fong:2021,Carlsten:2021}. Our results could help interpreting such measurements and provide testable predictions going to even lower mass regimes.

\begin{figure}
    \centering
    \includegraphics[width=\columnwidth]{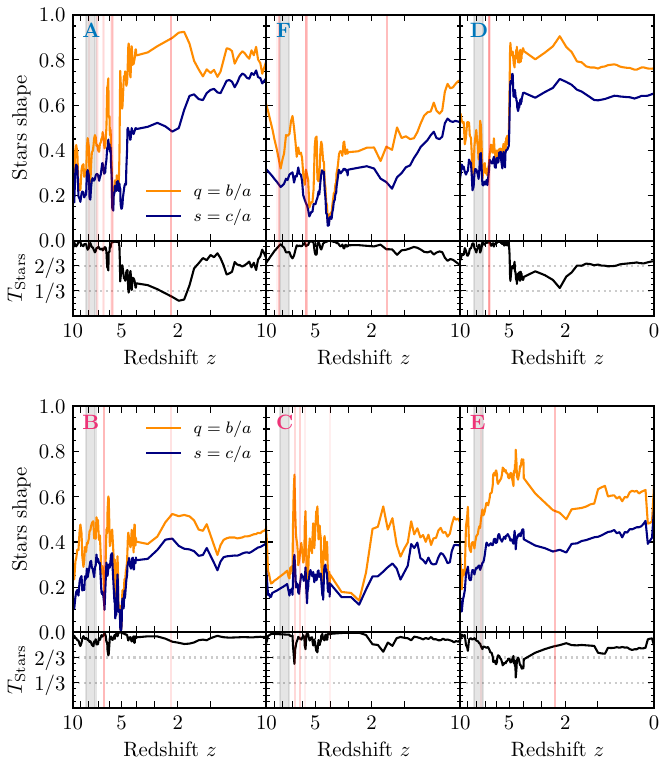}
    \caption{Evolution of the ellipsoidal axis ratios and triaxiality parameter for the stellar component of each simulated galaxy. The shapes are measured at fixed physical radius, corresponding to the $z=0$ stellar half mass radius, $r_{1/2}$. The top and bottom sets of panels correspond to the rejuvenated and relics galaxies, respectively. The orange and dark blue curves correspond to the intermediate-to-major (q) and minor-to-major (s) axis ratios, respectively.  Below each panel we show the evolution of the triaxiality parameter.}
    \label{fig:shape vs z stars}
\end{figure}

To better understand how the previous trend between shapes and star formation histories arise, we look into the evolution of the shape in these dwarfs. Fig. \ref{fig:shape vs z stars} presents the evolution of ellipsoidal axis ratios, $q$ and $s$, vs redshift for the main progenitors of the dwarf galaxies. The shapes are computed at fixed physical radius $r_{1/2} (z=0)$ at all times. This choice is made to avoid effects of the size evolution in the shape measurement. 
We find that rejuvenated galaxies (top row) and relics (bottom row) have similar axis ratios at early times, $z \gtrsim 7 $, with  $q \sim s \sim 0.2-0.4$. However, rejuvenators tend to increase their axis ratios and become more spherical at later times. The shape evolution appears to be strongly linked to the star formation activity of these dwarfs. In particular, for Halo A and D, the significant increase in the $q$ and $s$ values at $z\sim5$ coincides with the first peak of star formation episode after reionization, which for these galaxies corresponds to a significant sudden increase in stellar mass too (i.e. high SFR). Similarly, Halo F shows an increase in the axis ratios during its star formation period, although it is a gradual increase, reflecting the lower SFR and steady increase of stellar mass after $z\sim 1.5$. Moreover, the star formation in the rejuvenated dwarfs occurs most of the time in a spheroidal gas distribution rather than in a thin gas disc, which we suggest is the main mechanism imprinting the more spherical stellar distribution in the rejuvenated dwarfs. 

\begin{figure}
    \centering
    \includegraphics[width=\columnwidth]{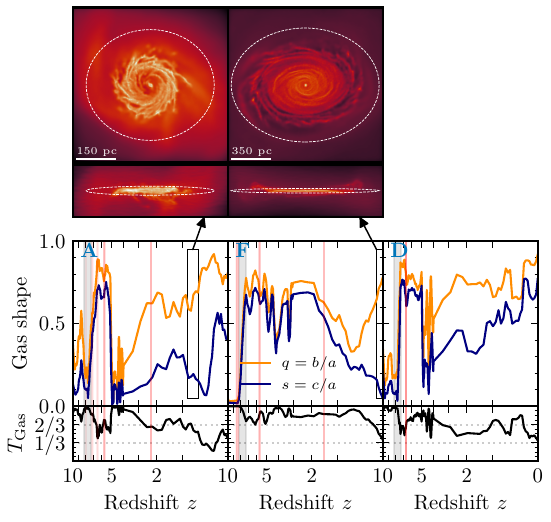}
    \caption{Evolution of the gas ellipsoidal axis ratios for the rejuvenated halos, A, F, and D. Shapes are measured in the same way as in Fig. \ref{fig:shape vs z stars} but for gas. The top panels show the gas surface density at times where disc-like distributions are detected (see Table \ref{tab:morphology-summary}). These are highlighted with a rectangular region on the bottom panels. }
    \label{fig:gas shape}
\end{figure}

Fig. \ref{fig:gas shape} shows the gas evolution across time for rejuvenated galaxies, considering all gas cells within $r_{1/2,\star}$, in the same way as with the stars
This highlights that in the most relevant star formation events, the gas distributions are usually spherical ($q \sim s \sim 1$). Additionally, in some of the halos, we are able to identify the presence of oblate gas distributions, i.e. gas discs. For these halos, in the top panels we visualise these discs with edge-on and face-on orientations, obtained from the eigenvectors of the inertia tensor (see Sec. \ref{sec: shape-algorithm}). Nevertheless, these are usually small and short-lived gas discs, and are not correlated with the bulk of star formation in these halos. Therefore, we argue that the birth stellar distribution, in combination with the relaxation during the evolution of the galaxies makes the rejuvenators evolve from prolate to triaxial or spheroidal distributions. 
On the other hand, relics do not show a significant change in their axis ratios at late times and tend to preserve their primordial prolate shape throughout most of their evolution.  

It is worth highlighting that the shape parameters for the stellar component (Fig. \ref{fig:shape vs z stars}) indicate that stars in the LYRA dwarfs are never dominated by a disc-like morphology ($s << 1$ and $q \sim 1$) in their evolution. They are born triaxial and stay triaxial or become spherical. 

\subsection{Influence of baryons on dark matter halos}
\label{sec:darkmatter}

\begin{figure}
    \centering
    \includegraphics[width=\columnwidth]{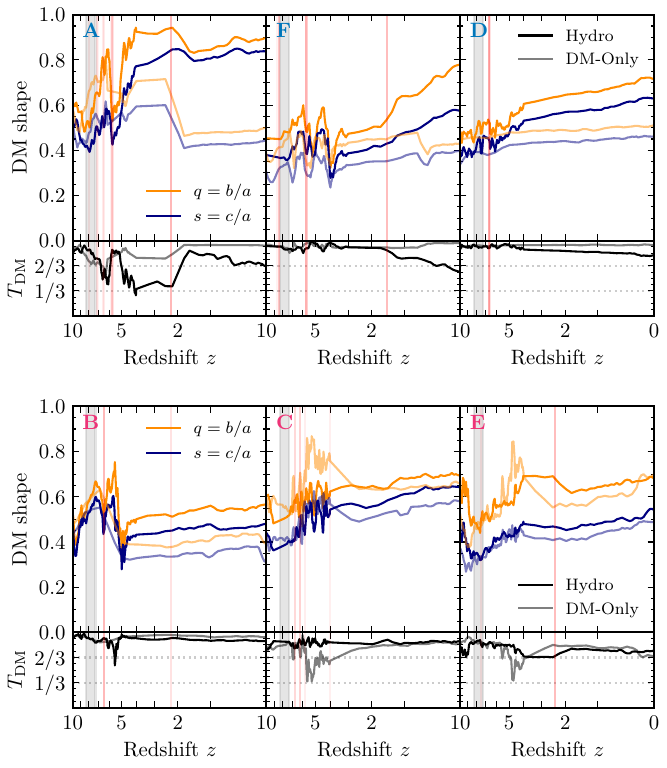}
    \caption{Similar to Fig. \ref{fig:shape vs z stars} but showing the evolution of the shape of DM components at fixed physical radius of $0.01R_{200\mathrm{c}} (z=0)$. The hydro and DMO-runs are shown as opaque and transparent lines respectively.}
    \label{fig:shape vs z dm}
\end{figure}

\subsubsection{The shape of the inner dark matter halo}

Our earlier results on the shapes of the stellar component motivate us to look into the shape of dark matter halos in our sample and its connection with SFHs. We compute the DM halo shapes of the six dwarfs, using the same method as we did for the stellar component, in both the hydrodynamical simulations and DMO runs. The redshift evolution of the inner DM halo axis ratios, measured at fixed physical radius of $0.01R_{200\mathrm{c}} (z=0) \sim 200-300$pc, are shown in  Fig. \ref{fig:shape vs z dm}, where lines with higher (lower) opacity correspond to hydro (DMO) runs. The choice of radius, $0.01R_{200\mathrm{c}} (z=0)$, is motivated by the stellar half mass radii which were used for measuring the shape of stars (see Table \ref{tab: dwarf info}). 
We note that DM distribution is systematically more spherical (i.e. larger $q$ and $s$ values) in all hydrodynamical runs compared to their counterparts in DMO runs, for most of the cosmic time. This is consistent with previous works reporting similar effects across different mass scales and at different galaxy formation models \citep[see e.g.,][]{Kazantzidis:2004,Vera-Ciro:2011,Orkney:2023}.

Moreover, we find that DMO runs are relatively similar across all halos with axis ratios in the range $\sim 0.4-0.6$, however, that is not the case in hydro runs. The halos of rejuvenator galaxies (top row) deviate from their DMO counterparts with time and are significantly more spherical at lower redshifts. In comparison, the halo shapes of relic galaxies stay closer to their DMO counterparts across all redshifts. We note that the redshift when hydro runs start to deviate from the DMO runs Fig. \ref{fig:shape vs z dm} correspond to the reignition of star formation post reionization. This is most clear in Halo A with axis ratios sharply increasing around $z \sim 5$. In Halo F and D, the axis ratios gradually increase during the star formation activity. Even the halo of the relic galaxy, Halo B, shows a small increase in its axis ratios coinciding with the short burst of star formation around $z\sim5$.    
Our results highlight the sensitivity of the shape of DM halos, in the inner regions, to the star formation activity of their central galaxies, in the regime of dwarf galaxies.   

\begin{figure}
    \centering
    \includegraphics[width=\columnwidth]{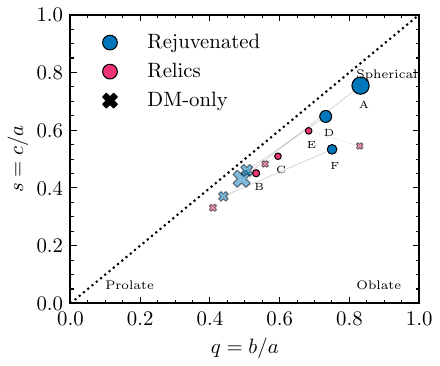}
    \caption{Axes ratios for the dark matter halo measured at a fixed radius of $r_{1/2,\star}$. The coloured points represent the rejuvenated and relic dwarfs in the hydro runs whereas the cross symbols show the corresponding DM-only counterparts. The symbol sizes are proportional to the fraction of stellar mass formed after reionization.} 
    \label{fig:DMO shapes}
\end{figure}

These results are summarized in Fig. \ref{fig:DMO shapes} where we show the axis ratios of DM halos at $z=0$, measured at $r_{1/2,\star}$. The hydro runs are displayed as circles, and are connected to their DM-only counterparts, shown as crosses, with a grey line. Similarly to Fig. \ref{fig: Stellar Axis ratios}, the halos are divided into two groups of rejuvenators (blue) relics (magenta), and the sizes of the symbols represent the stellar mass fraction formed after reionization. 
Considering that SFHs in the hydro runs are linked, to some degree, to the dark matter halo assembly and accretion history, one might wonder whether the shape of halos in DM-only runs show any correlation with the SFH activity in the counterpart hydro run. We see clearly in this figure that DM-only runs are clustered with no clear distinction between different SFH categories and the amount of star formation post reionization. Nevertheless, Halo E seems to be an outlier here, whereas we did not see any significant difference between this halo and the others in Fig. \ref{fig:shape vs z dm}. We checked that there is an ongoing merger that affects this measurement but if one measures closer to the centre, then the behaviour of the other relics is recovered. It should be noted that we measured the halo shape at $r_{1/2,\star}$, which is usually the radius that properly captures this effect. However, the trends discussed here might not be as clear when one measures the shape at a different radii (see Appendix \ref{app: DMhalo shape} and Fig. \ref{fig: app DM radial shape} for more details on this aspect).

In accordance with the results from Fig. \ref{fig:shape vs z dm}, we see that DM halo of rejuvenated galaxies are more spherical in the inner regions, than relics, in the hydro runs. Moreover, we see a correlation between the axis ratios and the amount of star formation post reionization.
In summary, we conclude that star formation histories of these dwarf galaxies affect the shape of their inner dark matter halos significantly.

Recently, \citet{Orkney:2023} carried out a thorough study of halo shapes in dwarf galaxies from the EDGE project, where a threshold in terms of halo mass ($M_{200\mathrm{c}} \leq 1.5 \times 10^9\, \mathrm{M}_{\sun}$) is proposed for the inner halo shape to be affected by the baryonic component of the galaxy. We do not find a similar dependence on the halo mass in our results. However, we note that the halo mass they propose divides their sample into rejuvenators and reionization relics\footnote{These are refereed to as reionization fossils in \citet{Orkney:2023}.}, such that higher mass halos host rejuvenators and these are the ones when the DM shape is affected. Therefore, our results are consistent with their findings, and we propose that the main factor affecting the DM halo shape is the extended star formation period post reionization.

\subsubsection{Halo density profiles}

\begin{figure*}
    \centering
    \includegraphics[width=2\columnwidth]{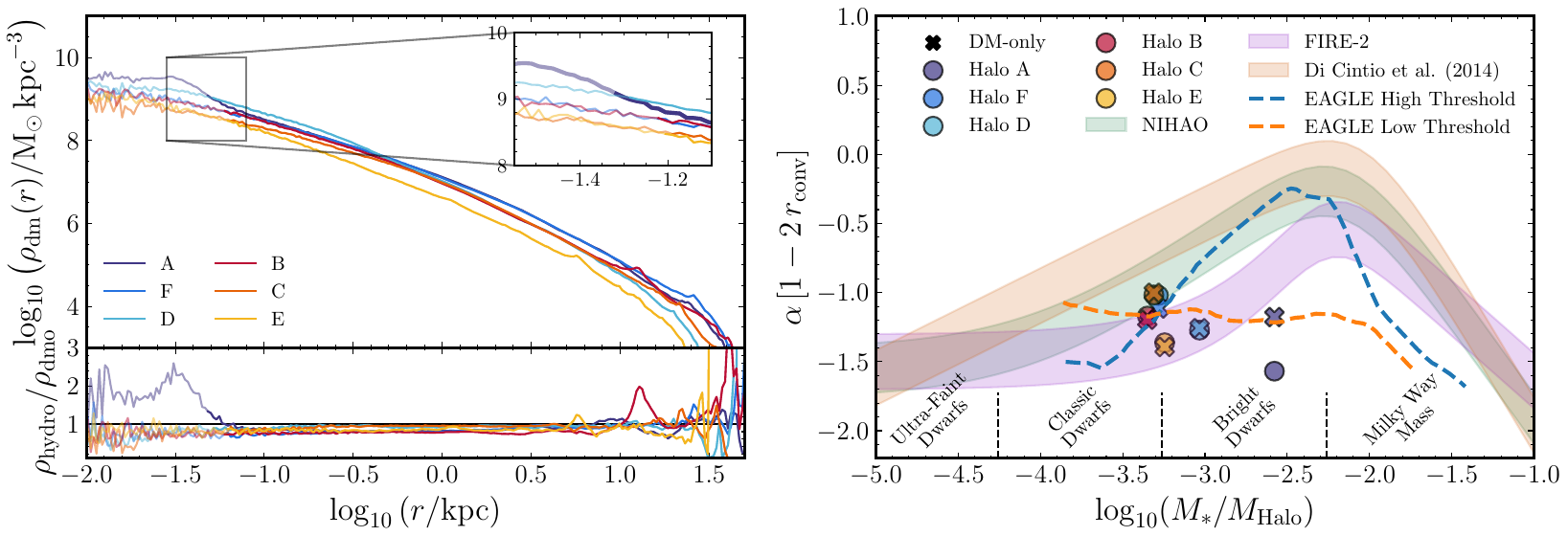}
    \caption{\textit{Left:} Dark matter density profiles for all halos. The panel at the bottom shows the ratio between the profile in the Hydro run over the one in the DM-only run. Lines are drawn fainter below the \citet{Power:2003} convergence criterion. The inset shows the region corresponding to the inner $100$ pc to highlight the contraction feature of Halo A. \textit{Right:} Inner slope of the density profile, measured between 1 and 2 times the convergence radius ($\sim 50-100$ pc), as a function of the stellar to halo mass ratio. For comparison, we include the predictions from other works such as \citep{DiCintio:2014}, the NIHAO simulations \citep{Tollet:2016}, FIRE-2 \citep{Lazar:2020} and the EAGLE model with high and low star formation threshold \citep{Benitez:2019}. From these two panels, we see that the inclusion of  baryons does not strongly affect the inner density slope, except for Halo A, which becomes denser towards the centre.}
    \label{fig:Density profiles}
\end{figure*}

The influence of extended star formation period on the shape of DM halos in LYRA dwarfs raises the question on whether the central DM density profiles are affected, as suggested by other works \citep{Navarro:1996b,Read:2005,Pontzen:2012,Chan:2015-FIRE,Benitez:2019}. To address this question, we look into the DM density profiles of LYRA dwarf galaxies in the left hand side of Fig \ref{fig:Density profiles}, where the top panel shows the $z=0$ profiles, computed in spherical apertures, for the six hydro runs, and the bottom panel shows the ratios relative to their DM-only counterparts. The profiles are shown with a fainter line below the convergence limit of \citet{Power:2003}, applied to the DM particles only. We see no central flattening of dark matter profiles in the hydro runs. All density profiles in the hydro runs follow their DM-only counterparts closely\footnote{Differences in the outskirts, $r > 10$ kpc, are due to individual subhalos and not relevant to the discussion here.}, with the exception of Halo A that surprisingly exhibits a steepening of the profile and increased dark matter towards the centre of the halo.
We will return to the case of Halo A in the next subsection.

It is argued in the literature that the impact of baryons on the dark matter density profiles correlates with the stellar to halo mass content of a galaxy \citep{DiCintio:2014,Tollet:2016}. We therefore investigate our results in the light of those findings in the right hand panel of Fig. \ref{fig:Density profiles} where we show the inner slope of the DM density profiles in our hydro runs (circles) and DM-only counterparts (crosses) as a function of the stellar-to-halo mass ratios in the hydro runs. The inner slope, is computed by assuming a power law profile, $\rho_\mathrm{DM} \propto r^{\alpha}$ and measured in the range $(1-2) \times r_{\rm conv}$. \footnote{We select the larger convergence radius, $r_{\rm conv}$, between the hydro and DM-only runs. This is to ensure that for both runs the slopes are computed within the convergence limit.} This effectively corresponds to a physical radius of $\sim 50-100$ pc from the centre of each galaxy. As expected from the previous discussion, we see that our hydro runs and their DM-only counterparts are on top of each other, except for Halo A, where the steepening of the inner DM profile in the hydro run is seen as a lower $\alpha$ value    
For reference, we also show the predictions from \citet{DiCintio:2014, Tollet:2016, Lazar:2020} and \citet{Benitez:2019}, who discuss the formation of dark matter density cores in different simulations, NIHAO, FIRE-2, and EAGLE \footnote{Note that all these works measure the slope at $1.5\%\, r_\mathrm{vir}$, which is too large for the LYRA galaxies as we want to measure the innermost region of the density profile ($\sim 50$ pc). In Fig. \ref{fig:slopes appendix} we show the slope measured at $1.5\%\, r_\mathrm{vir}$ for comparison.}. For the EAGLE model, we show both the low and high threshold for star-formation discussed in \citet{Benitez:2019}. We see that the stellar-to-halo mass ratio of most LYRA galaxies lie in the range where other models predict little to no change in the inner slope of DM densities. Hence it is not surprising that we do not see any cores in the LYRA dwarfs, at $z=0$. The only halo with a stellar-to-halo mass ratio in the expected range for the formation of dark matter density cores is Halo A, which is the halo with overdensity of DM in the centre. 

We have inspected the evolution of the density profiles of our sample over time to check whether they ever formed dark matter density cores in their evolution. We find that the previous conclusions hold across time; i.e., the central dark matter content in the hydro runs follow the DM-only ones across time, except for Halo A. In order to better understand the reason, we look at the evolution of gas content in these dwarfs. Previous studies postulated that the formation of dark matter core is linked to fluctuation in the gravitational potential due to rapid changes in the gas content in the central region as a result of supernovae feedback \citep{Navarro:1996b,Read:2005,Pontzen:2012}. \citet{Benitez:2019} shows that this process is only efficient when the gas can dominate the mass in the central regions, and the subsequent removal of this gas by the SN feedback, producing large enough changes in the gravitational potential \citep[see also ][]{DiCintio:2014,Chan:2015-FIRE,El-Zant:2016,Tollet:2016,Wetzel:2016-FIRE,Fitts:2017-FIRE}.

\begin{figure}
    \centering
    \includegraphics[width=\columnwidth]{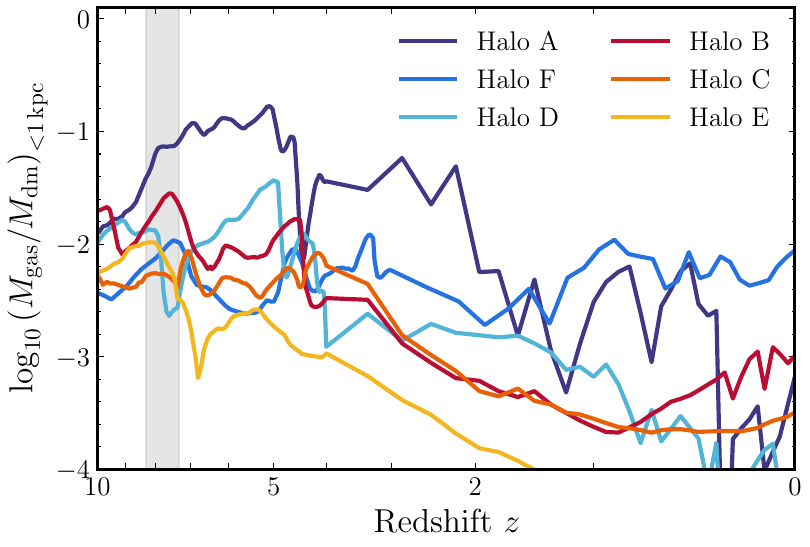}
    \caption{Evolution of total gas mass to DM mass ratio within 1 kpc for all simulated dwarfs. This shows that gas never dominates the mass budget in any of the galaxies, hence explaining the lack of a dark matter core.}
    \label{fig:fgas}
\end{figure}

In Fig. \ref{fig:fgas}, we show the evolution of the gas fraction (relative to the DM), $M_{\rm gas}/M_{\rm DM}$, inside 1kpc for the main progenitor of LYRA dwarfs as a function of redshift. This figure shows that the gas fraction in our sample never exceeds 10\% in the inner regions, and it is typically below 1\%. These values are more in par with gas fractions in APOSTLE and Auriga dwarf galaxies which also do not show any sign of DM core formation, as shown in \cite{Bose:2019}. \citet{Benitez:2019} shows that the gas fractions, hence DM core formation, can increase in these models when the star formation threshold increases allowing the gas to become denser in the inner regions. We emphasize that mass resolution in the LYRA simulations is 4 orders of magnitude higher than ones discussed \citet{Benitez:2019} and \cite{Bose:2019}. Hence the arguments about star formation threshold does not apply here.  We argue that the low mass of our haloes, which overall translates into low baryonic fraction, is the main reason that our dwarf galaxies do not have any DM cores throughout their life. 

Recently, \citet{Muni:2025} presented a study that relates baryon-driven dark matter core formation in the EDGE simulations \citep{Agertz:2020, Rey:2022} with the ratio of post and pre reionization stellar mass, $M_{\star,\rm post}/M_{\star,\rm pre}$. They highlight that this evidences the importance of the timing of star formation, with respect to reionization, providing even tighter constraints than with the stellar to halo mass ratio. Their findings show that haloes transition from a classical NFW profile to a cored one as $M_{\star,\rm post}/M_{\star,\rm pre}$ increases. While LYRA galaxies do not form dark matter cores, the entire sample have small $M_{\star,\rm post}/M_{\star,\rm pre}$ values. All of them, but Halo A ($M_{\star,\rm post}/M_{\star,\rm pre} = 1,44$), have ratios smaller than $1$. In principle, this is in agreement with the findings of \citet{Muni:2025}, although we need a larger sample, including core forming halos to provide conclusive interpretations about this. We also stress that the redshift used for reionization between LYRA and EDGE differ, hence we are separating our populations at different times.

\subsubsection{Curious case of Halo A}
\label{sec:case Halo A}

In all our previous results, Halo A, stands out as an unusual case. In particular, it shows the most significant change when comparing the dark matter halo properties in the hydro vs the DM-only run. This is especially noticeable in the dark matter density profile, where we find a steepening (or overdensity) towards the centre in the hydro run (see Fig. \ref{fig:Density profiles}). We investigate the evolution of halo A in more details in this subsection.

\begin{figure}
    \centering
    \includegraphics[width=\columnwidth]{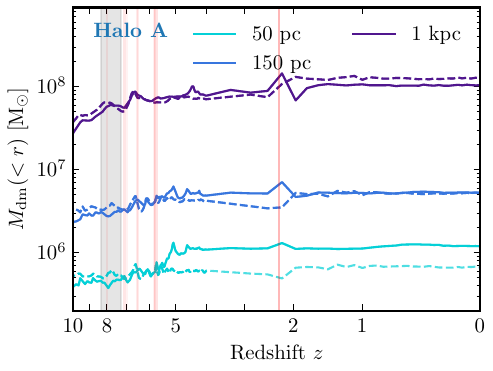}
    \caption{Dark matter mass evolution at different radii for Halo A. Solid lines show the full-hydro run and dashed lines show the corresponding dark matter only run. The gray vertical region illustrates the time of reionization and the red vertical lines correspond to mergers. From this figure, we highlight the factor $\sim 2$ increase in the enclosed dark matter mass in the inner 50 pc. This is only present in the full-hydro run and hence it is attributed to the effect of baryons.} 
    \label{fig: Halo A DM Growth}
\end{figure}

We show the evolution of enclosed dark matter mass in various radii of halo A in Fig. \ref{fig: Halo A DM Growth}, for the hydro and the DM-only runs, as solid and dashed lines, respectively. There is a noticeable sudden increase of the dark matter content in the inner 50 pc of Halo A occurring at $z \sim 5$. By comparing the timing of this increase and the baryonic evolution in Fig. \ref{fig:Mass growth and SFH}, we can infer that the halo contraction is connected to the star formation event with a high SFR around $z \sim 5$. 


\begin{figure}
    \centering
    \includegraphics[width=\columnwidth]{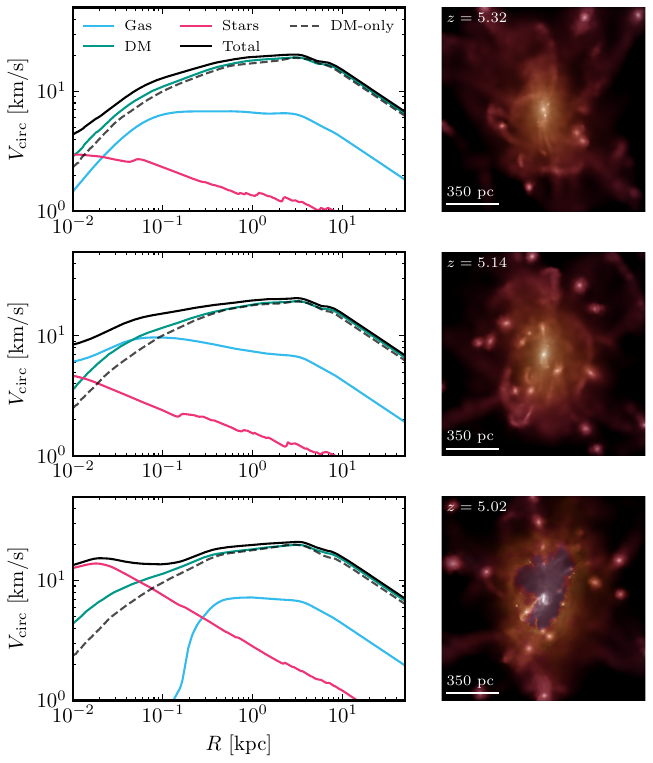}
    \caption{From top to bottom the panels show the circular velocity profile (left) and surface density of stars and gas (right) of Halo A on different snapshots during the starburst at $z\sim5$. This illustrates the contraction of the dark matter halo being attributed to the condensation of the gas in the centre followed by a strong star formation event.}
    \label{fig: Halo A Contraction}
\end{figure}

We explore the burst of star formation and the change of the DM distribution more closely in Fig. \ref{fig: Halo A Contraction}, where we show the circular velocity profiles for different components of Halo A in the middle column. We present three different redshifts in different rows, corresponding to before, at the start of and shortly after the star formation burst ($z\sim 5$). We see that the potential is dominated by the DM at all radii in the earlier snapshot (first row) and the profile in the hydro run is closely following the DM-only counterpart. At the start of the star formation (middle row), the gas has become much denser in the central regions, even dominating the mass within $\sim 30$ pc. Shortly after the star formation begins (bottom row), the gas is quickly removed from the inner regions due to SN feedback, but the central mass continues to be dominated by baryons, and this time it is the newly born stars. 
The enhanced baryonic mass towards the centre of the galaxy deepens its gravitational potential, causing the DM to become denser towards the centre, explaining the contraction of the DM halo lasting until $z=0$. The absence of a DM-core can then be understood as a consequence of the stars dominating the gravitational potential before the feedback quickly removes the gas from the centre.
We note that the central DM over density in Halo A is mostly below the convergence radius (see Fig. \ref{fig:Density profiles}), hence the detail of its structure in affected by numerical resolution. However the inset in Fig. \ref{fig:Density profiles} highlights the steepening at radii just above the convergence limit.  

We emphasize that earlier works that studied the contraction of dark matter halos and its relation with the baryonic component of galaxies \citep{Blumenthal:1986, Duffy:2010, Dutton:2016, Cautun:2020}, were carried out on a different mass scale, mostly on the Milky Way mass regime. The difference in the  dwarf galaxy regime is that they are mostly dark matter dominated and lose the majority of their gas. Therefore, the halo contraction effect is, on average, not expected for dwarfs and what we observe for Halo A is not common at the mass scale relevant to our study.

\section{Conclusion}\label{sec:conclusions}

In this work, we investigate the evolution of the baryonic and dark matter content of six simulated dwarf galaxies from the LYRA project \citep{Gutcke:2022-LYRAIII}. These dwarf galaxies have a small range of stellar mass, $M_\star=4\times10^5-10^7 \msun$, that puts them amongst lower mass end of the classical dwarf regime, with their halo masses in the range $7\times10^8-5\times10^9 \msun$. According to their star formation histories (SFH), these dwarfs can be divided into two distinct categories, reionization relics and rejuvenators. Reionisation relics show very little to no star formation post-reionization, whereas rejuvenators have sustained period of star formation. This can be quantified in terms of the fraction of stars born post-reionization, i.e. $f^{\star}_{\rm post-reio}=M_\star (< z_{\rm re})/M_\star({z_0})$.  
We look into the effect of such distinct SFHs on properties of the galaxies and their dark matter haloes. The main conclusions of this work are as follows:

\begin{itemize}

    \item The SFHs leave their signatures on the metallicity ($[\rm{Fe/H}]$) distribution of these dwarfs, by changing the shape, as well as the median (and peak) of the distributions. Relics, which show little or no star formation after reionization, have a prominent tail of low metallicity stars and their median $[\rm{Fe/H}]$ is lower (Fig. \ref{fig:Metallicity}). We indeed see a correlation between $f^{\star}_{\rm post-reio}$ and median $[\rm{Fe/H}]$. Our results indicate that the scatter in the mass-metallicity relation correlates with $f^{\star}_{\rm post-reio}$.        
    
    \item By measuring the intrinsic shape of these dwarf galaxies, using the principle axis ratios of the stars inside $r_{1/2, \star}$ (Fig. \ref{fig: Stellar Axis ratios}), we find that rejuvenated dwarfs are more spherical compared to the relics, with a correlation between axis ratios and $f^{\star}_{\rm post-reio}$. 

    \item Relics and rejuvenators have roughly similar triaxial/prolate shapes until reionization and shortly after (see Fig. \ref{fig:shape vs z stars}). Reionization relics stay close to their primordial prolate morphology during their lifetime. In particular, after a dwarf gets quenched due to reionization, its morphology does not evolve significantly afterwards. On the other hand, the rejuvenators evolve to become more spherical as they form more stars post reionization. We see a sudden increase in axis ratios (getting more spherical) following bursts of star formation post-reionization in some cases.
    
    \item We attribute the previous findings to the significant role of mergers in the stellar assembly of dwarf galaxies prior to reionization resulting in a prolate/triaxial shape at early time, followed by the formation of in-situ stars post-reionization from mostly spheroidal gas distribution, in the case of rejuvenators. The gas component in this later dwarf galaxies are not dominated by a disc-like component majority of the times.

     \item The shape of the dark matter haloes in the inner regions ($<0.01R_{200}$) are rounder in the hydro runs compared to their DMO counterparts. This difference is relatively smaller for relics and the hydro and DMO runs follow each other closely over time (Fig. \ref{fig:shape vs z dm}). However, the difference between hydro and DMO runs is much more pronounced for the rejuvenators. They become increasingly rounder with more star formation.  

    \item Despite the effect of baryons on the shape of the DM halos in the inner regions in our simulated dwarf galaxies, their dark matter density profiles are mostly unaffected compared to their corresponding DM-only runs and follow cuspy NFW profiles, without ever forming DM cores.
    Nevertheless, one of the halos gets contracted in the full-hydro run (\S \ref{sec:case Halo A}). A detailed analysis of the reasons for this contraction points towards the effect of the adiabatic contraction of the gaseous component, followed up by a significant star formation event that makes the gravitational potential to be dominated by baryons, even after the gas removal by the supernovae events.

\end{itemize}

In this analysis, we have not distinguished between in-situ and ex-situ star formation. The SFHs are built from the formation time of all stars bound to any given host at $z=0$. We find that star formation activity post-reionization is largely dominated by in-situ star formation. However, mergers play a non-negligible role in the evolution of the galaxies in the same period. For example, significant mergers occurring post reionization ($z<7.3$) contribute $15-40$ percent to the final stellar mass of the galaxies. Despite the lack of star formation post-reionization, the reionization relics acquire stellar mass through mergers post-reionization. We also find, in some cases, periods of increased SFR following significant mergers, or change in the intrinsic shape of stars and/or dark matter following the mergers. The role of mergers in the evolution of dwarf galaxies is much debated in the literature and is an interesting topics, which we will follow up in future works.

Our results highlight the diversity in the properties of dwarf galaxies with $M_{\rm halo} \sim 10^9 \msun$ which can be traced back to the interplay between their halo assembly histories and reionization, and the subsequent ability of halos to form star after reionization. Comparing our results with other high resolution simulations and strengthening our conclusions using larger number of simulated galaxies will be of importance as a next step.
The correlations predicted in our study between the star formation histories, metallicity and shape of low mass isolated dwarfs galaxies can be tested by upcoming data from surveys such as Rubin or Euclid which will discover and measure the properties of a large population of dwarf galaxies around the Local Group in relative isolation.  

\section*{Acknowledgements}

J.S. acknowledges support from the Science and Technologies Facilities Council (STFC) through the studentship grant ST/X508354/1, and the support from an ESO Studentship. S.T.B. and A.F. have been supported by a UK Research and Innovation (UKRI) Future Leaders Fellowship [grant no MR/T042362/1] and a Sweden's Wallenberg Academy Fellowship. S.B. is supported by the UKRI Future Leaders Fellowship [grant numbers MR/V023381/1 and UKRI2044]. J.E.D is supported by the United Kingdom Research and Innovation (UKRI) Future Leaders Fellowship `Using Cosmic Beasts to uncover the Nature of Dark Matter' (grant number MR/X006069/1). This work used the DiRAC@Durham facility managed by the Institute for Computational Cosmology on behalf of the STFC DiRAC HPC Facility (www.dirac.ac.uk). The equipment was funded by BEIS capital funding via STFC capital grants ST/K00042X/1, ST/P002293/1, ST/R002371/1 and ST/S002502/1, Durham University and STFC operations grant ST/R000832/1. DiRAC is part of the National e-Infrastructure.

\section*{Data Availability}

The data presented here is available upon reasonable request to the corresponding author.



\bibliographystyle{mnras}
\bibliography{example} 




\appendix

\section{Ellipsoidal shape of the Gas component}\label{app: gas shape}

\begin{figure}
    \centering
    \includegraphics[width=\linewidth]{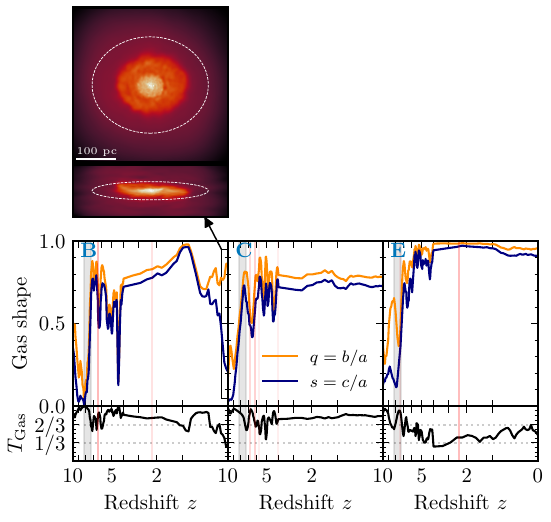}
    \caption{Evolution of the ellipsoidal axis ratios for the gas component for the relic dwarfs in the LYRA sample. }
    \label{fig: app gas-evo}
\end{figure}

For completeness, in Fig. \ref{fig: app gas-evo} we show the evolution of the ellipsoidal axis ratios, $q$ and $s$ for the gas component for Halos B, C and E, i.e. reionization relics. To compute the inertia tensor and obtain the shapes, we are using all the gas within the galaxy, with no distinction between cold and hot gas. We stress that although these halos have little to none cold gas (see. Fig. \ref{fig:Mass growth and SFH}), they still retain some hot and diffuse gas that allows to compute the inertia tensor. This shows that relic galaxies tend to have spherical gas distributions, which remain largely unperturbed due to the lack of SNe feedback. The most interesting case is Halo B, that shows hints of a disc like component forming at late times, which coincides with the small star formation episode of this halo. However, this is a very small feature, as shown by the scale of the visualisation on top of the Halo B panel.

\section{Present-day Dark Matter halo shape}\label{app: DMhalo shape}

Since the shape of the dark matter halo depends on the chosen radius to measure it, we show in Fig \ref{app: DMhalo shape} the shape of the $z=0$ LYRA dark matter halos. We see the influence of substructure, especially at larger radius. The general trend is that the shape gets rounder as we move to the centre and it is more prominent on the rejuvenated dwarfs.

\begin{figure*}
    \centering
    \includegraphics[width=\textwidth]{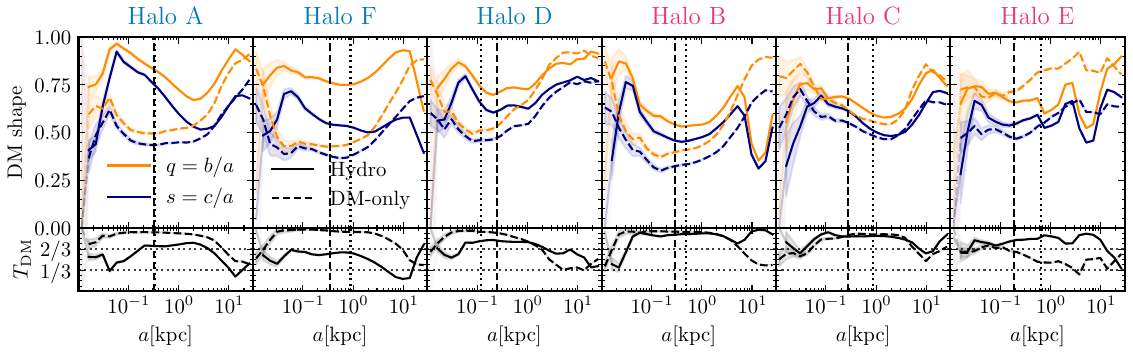}
    \caption{DM halo axis ratios ($q$ and $s$) against the major axis of the ellipsoid for all LYRA halos at $z=0$. Solid lines correspond to the hydro runs, whereas dashed lines show the DM-only runs. Vertical dashed and dotted lines represent $0.01R_{200\rm c}$ and $r_{1/2}^\star$ for each halo respectively. Bottom panels show the Triaxiality parameter (Eq. \eqref{eq: Triaxiality})}
    \label{fig: app DM radial shape}
\end{figure*}

\section{Slope of the density profiles within 1 percent of the virial radius}\label{app: slope at 1percent rvir}

\begin{figure}
    \centering
    \includegraphics[width=\linewidth]{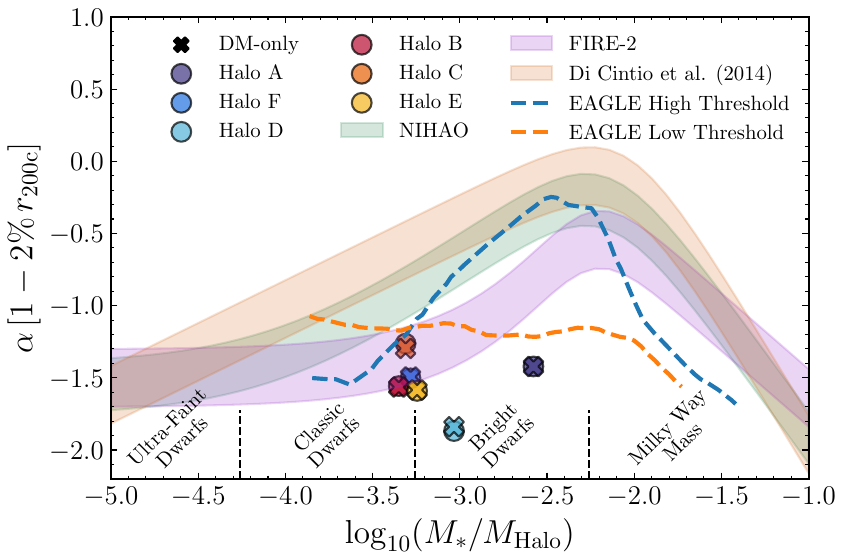}
    \caption{Same as right panel of Fig. \ref{fig:Density profiles} but measured between 1 and $2\% \, r_\mathrm{vir}$. This shows there is virtually no difference in the dark matter density profiles between the hydro and the DMO runs at these radii.}
    \label{fig:slopes appendix}
\end{figure}

To have a proper comparison with the literature on the measurement of the slope of the dark matter density profile, we include the inner slope for all LYRA halos measured between 1 and $2\% \, r_\mathrm{vir}$, in Fig. \ref{fig:slopes appendix}

Importantly, in this regime, we see no significant impact of baryons on the slope of the density profile. We argue this is due to the small size of the LYRA galaxies that makes any effect to be located in the innermost region of the halos, well within $1\%\,r_\mathrm{vir}$.


\bsp	
\label{lastpage}
\end{document}